\documentclass[
 aip,
amsmath,amssymb,
reprint,
]{revtex4-1}

\usepackage{xcolor}
\usepackage{graphicx}
\usepackage{dcolumn}

\renewcommand{\=}{\mbox{\,=\,}}
\newcommand{\e}{\mathrm{e}}

\begin{document}

\title{Controlling thermodynamics of a quantum heat engine with modulated amplitude drivings}

\author{Sajal Kumar Giri}
\affiliation{Department of Chemistry, Northwestern University, 2145 Sheridan Rd., Evanston, IL 60208, United States}
\author{Himangshu Prabal Goswami}
\email{hpg@gauhati.ac.in}
\affiliation{Department of Chemistry, Gauhati University, Jalukbari, Guwahati-781014, Assam, India}
\date{\today}

\begin{abstract} 
External driving of bath temperatures with a phase difference of a nonequilibrium quantum engine leads to the emergence of geometric effects on the thermodynamics. 
In this work, we modulate the amplitude of the external driving protocols by introducing envelope functions and study the role of geometric effects on the flux, noise and efficiency of a four-level driven quantum heat engine coupled with two thermal baths and a unimodal cavity.
We observe that having a finite width of the modulation envelope introduces an additional control knob for studying the thermodynamics in the adiabatic limit.
The optimization of the flux as well as the noise with respect to thermally induced quantum coherences becomes possible in presence of geometric effects, which is hitherto not possible with sinusoidal driving without an envelope. 
We also report the deviation of the slope and generation of an intercept in the standard expression for efficiency at maximum power as a function of Carnot efficiency in presence of geometric effects under the amplitude modulation. 
Further, a recently developed universal bound on the efficiency obtained from thermodynamic uncertainty relation is shown not to hold when a small width of the modulation envelope along with a large value of cavity temperature is maintained.
\end{abstract}

\maketitle

\section{Introduction}
Quantum heat engines (QHEs) have come a long way from the theoretically predicted Schulz-duBois engine \cite{PhysRevLett.2.262} to experimentally realizable engines. 
Notable examples include Rb based cold atomic setup \cite{PhysRevLett.119.050602}, Li-based Fermi gas \cite{science.342.713}, diamond based N-vacancy centres \cite{PhysRevLett.122.110601},
Paul-trapped Yb and Ca ion setups \cite{NatComm.10.1, science.352.325} and utilizing proton’s nuclear spin dissolved in 13-C labeled CHCl$_3$ \cite{PhysRevLett.123.240601}.
Role of coherences on the quantum thermodynamic and transport properties, establishing the validity of nonequilibrium fluctuation theorems, thermodynamic uncertainity relationships (TUR) are now being investigated experimentally and compared with the results obtained from several theoretically established models \cite{arXiv.2201.01740, PhysRep.694.1, PhysRevA.100.042119, PhysRevRes.2.023245}. 
Most of the theories are based on Markovian master equations and have seemed to agree pretty well with experimental observations \cite{PhysRevA.100.042119, NewJPhys.23.065004}. 
Success of such master equations in understanding several steadystate properties of QHEs led to the widespread use of another class of master equations that theoretically predict dynamics of quantum systems where system parameters are modulated in time, usually called driven dynamics \cite{arXiv.2202.06651, PhysRevLett.119.170602, PhysRevLett.127.200602}. 
Toy models based on QHEs are often a common choice to study driven dynamics using adiabatic master equations 
\cite{PhysRevE.99.032108, arXiv.2202.06651}.  
In such driven systems, periodic or nonperiodic modulation of a system parameter (like energy, reservoir temperature etc.) in an adiabatic fashion \cite{takahashi2020full,NewJPhys.20.113038, PhysRevB.102.155407,eglinton2022geometric,scopa2018lindblad} has led to the theoretical prediction of exotic properties such as creating new phases of matter and loss of tunneling which are corroborated using Floquet theory coupled to adiabatic master equations\cite{NewJPhys.14.123016, ye2021floquet,NewJPhys.20.053063,dann2018time,scopa2018lindblad}.
Further, adiabatic master equations developed by modulating two system parameters have been shown to break nonequilibrium fluctuation theorems and TUR because of the emergence of geometric phaselike quantities \cite{wang2022geometric, PhysRevE.96.052129, PhysRevB.102.245420,takahashi2020full,PhysRevLett.104.170601}. 
Although driven QHEs (dQHEs) have not yet been  experimentally realized, driven molecular junctions (theory of which is akin to QHEs) have been experimentally studied where geometric phaselike effects were proven to exhibit nonstandard influence on transport properties as predicted by adiabatic master equations \cite{JPhysChemC.122.1422, PhysRevB.93.195441}.  
With the current experimental realization of QHEs and driven molecular junctions, it is not far that, driven dynamics predicted by adiabatic master equations can be soon compared with experimental results. \\

There are several ways of driving the internal parameters of a dQHE. 
A particular example includes a stepwise sweep of the temperatures of the thermal reservoirs \cite{PhysRevLett.126.210603}. 
Such periodic driving protocols have led to the development of a quantum version of TUR signifying a trade-off between entropy production rate and signal to noise ratio\cite{PhysRevLett.126.210603}. 
Interestingly, over the past couple of years, several TUR have been developed in quantum engines\cite{NatPhys.16.15, JPhysA.54.314002, PhysRevLett.125.260604, PhysRevLett.126.010602}.  
In a previous study we have showed that, such a trade-off is invalid in presence of continuous driving of the temperatures of the two baths in a sinusoidal manner\cite{PhysRevE.96.052129}. 
We have also showed how other thermodynamic quantities of a popular QHE model such as flux, noise, efficiency, power etc. are influenced by such drivings \cite{PhysRevE.96.052129, PhysRevE.99.022104}.  
Notably, we  showed the universal linear slope of $1/2$ in the standard efficiency at maximum power (EMP) as a function of Carnot efficiency ($\eta_{c}$) no longer holds when there is a finite phase difference between the two continuous driving protocols\cite{PhysRevE.96.052129}.  
A natural question is how would the thermodynamic quantities behave when the continuous driving is replaced by an amplitude modulated driving (similar to a single cycle pulse).  
To keep things simple, we first focus on the adiabatic limit, where there are no sudden modulation or pulse induced dynamics in the engine, i.e, the driving timescale is well separated from the engine-evolution timescale. 
By considering two type of envelope functions Gaussian and Lorentzian, we note some interesting observations and compare the results obtained with the known continuous sinusoidal driving which is the limiting case with large envelope width. \\ 

This paper is organized as follows. 
In Sec.\ref{model}, we briefly introduce the model and discuss the basic underlying principles. 
In Sec.\ref{results}, we present our results and offer a discussion followed by the concluding remarks in Sec.\ref{conc}.  \\

\section{Amplitude modulated driven quantum heat engine}
\label{model}

\begin{figure}
    \centering
    \includegraphics[width = 0.5 \textwidth]{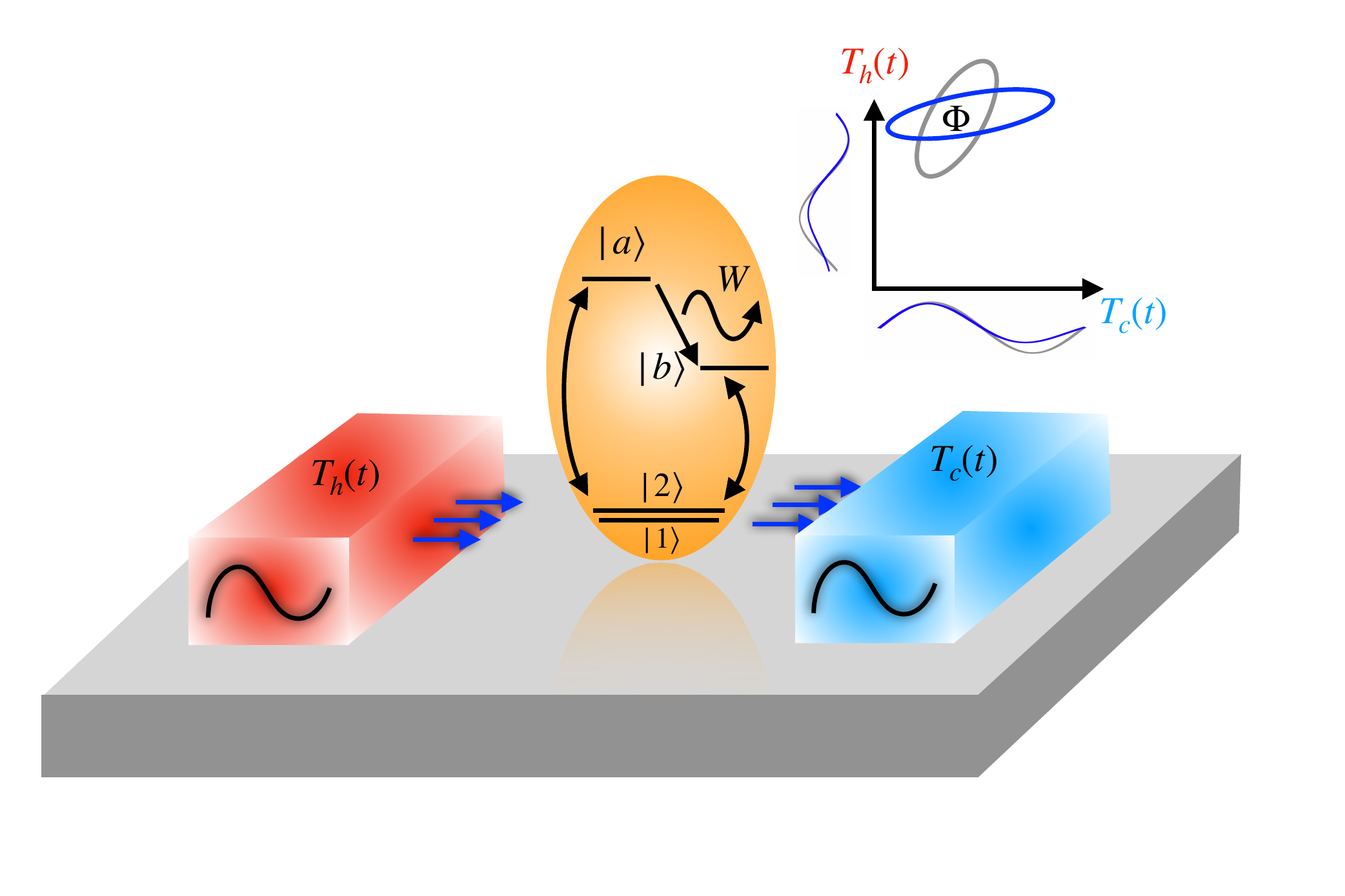}
    \caption{Schematic plot of an amplitude modulated driven $4$ level QHE. 
    Two degenerate states $|1\rangle$ and $|2\rangle$ are coupled with higher energy states $|a\rangle$ and $|b\rangle$ through respective thermal baths. 
    The state $|a\rangle$ is higher in energy than the state $|b\rangle$. Hot and cold bath temperatures are labeled as $T_{h}(t)$ and $T_{c}(t)$ respectively. 
    States $|a\rangle$ and $|b\rangle$ are also coupled with a unimodal cavity. 
    During the transition from $|a\rangle$ to $|b\rangle$ one photon is produced in the cavity with an energy equal to the energy difference between these states which we treat as work done by the system denoted by $W$. 
    Temperature amplitude shaping modifies the induced geometric phase $\Phi$ shown in the upper inset.}
    \label{qhe_fig}
\end{figure}

We consider a four level temperature driven quantum heat engine coupled with two thermal baths and a unimodal cavity, Fig.(\ref{qhe_fig}).
This model has been studied in several previous works \cite{PNAS.108.15097, PhysRevA.88.013842, PhysRevA.86.043843, EuroPhysLett.99.50005, PhysRevE.96.052129}. 
The theoretical framework has already been developed and discussed before \cite{PhysRevE.96.052129, PhysRevE.99.022104} and we refer to the appendix for necessary details. 
The engine operates in such a way that two thermal baths at temperatures $T_{h}(t)$ and $T_{c}(t)$, are adiabatically driven externally. 
The driving protocol is cyclic whose amplitude is being modulated in time thus shaping the envelope, which we refer to as {\em amplitude modulation}. 
We choose the following driving protocols, 
\begin{align}
     \label{temp_eq}
     T_{c}(t)&\=T_{c0}+A_{i}(t)\sin(\omega t), \\ 
     T_{h}(t)&\=T_{h0}+A_{i}(t)\sin(\omega t + \phi),
\end{align}
where $A_{i}$ is expressed as
\begin{align}
    \label{sin_env_eq}
     A_{S}(t)&\= A_{0}, \\
    \label{gaussian_env_eq}
     A_{G}(t)&\=A_{0}\exp\Bigg(-4\ln2\frac{t^{2}}{t_{e}^2}\Bigg),\\
    \label{lor_env_eq}
     A_{L}(t)&\=A_{0}\frac{[t_{e}/2]^2}{t^{2}+[t_{e}/2]^2}, 
\end{align}
with $i\in(S, G, L)$.
Here $t_{e}$ is termed as envelope duration and $A_{i}(t)$ is the envelope type such that the subscript $i$ represents the type of envelope -- constant, Gaussian, or Lorentzian. 
Note that $t_{e}$ is the full-width at half maximum (FWHM) for both Gaussian and Lorentzian envelopes. 
$A_0$, $\omega$ and $\phi$ are amplitude, frequency and phase difference between the driving protocols respectively. 
Here the cold (hot) bath temperature oscillates around $T_{c0}$($T_{h0}$). 
Bath temperatures are periodically driven in time such that $T_{h}(t)>T_{c}(t)$ condition is maintained throughout. Note that, the geometric contributions get explicitly added to the engine's thermodynamic properties due to the periodic driving of the reservoir temperatures. 
It is finite only when the driving protocols are phase different (which is introduced as a phase difference $\phi$) \cite{PhysRevE.99.022104}. 
Although we can observe driven dynamics when $\phi\,=\,0$, geometric contributions change the driven dynamics if and only if $\phi\,\ne\,0$. 
In this QHE, the exact analytical nature of the relationship between geometric effects and $\phi$ is however not known and so we resort to numerics to gain insights on its role on the thermodynamics.  
Throughout the text, whenever we refer to the phrase {\em `in the presence of geometric contributions'}, we mean $\phi\,\ne\,0$ in the driving protocols. 
The central quantity of interest in this work is the effect of geometric contributions  on the thermodynamics of the QHE. 
%Whenever we refer to the phrase {\em  `dynamic contributions}, we simply mean $\phi=0$, in the driving protocols.
The work done by the engine $W$ is quantified as energy flow (in the form of photon) into the cavity during the transition from $|a\rangle$ to $|b\rangle$. 
The hot and cold reservoirs induce coherence in the reduced system density matrix and they are denoted as $p_{h}$ and $p_{c}$ respectively \cite{CohOptPhe.1.7}. 
Through the amplitude modulation in the driving protocols, the additional parameter FWHM (or envelope duration), $t_{e}$, allows us to control the overall geometric contributions to the thermodynamics of the QHE. 
In the next sections, we focus on the thermodynamic quantities as a function of the control parameters, viz. 
envelope duration, $t_{e}$ and the hot bath induced coherence parameter, $p_{h}$. \\  

\section{Results and Discussion}
\label{results}

\subsection{Flux and Noise}
\label{flux_noise}

\begin{figure}
    \centering
    \includegraphics[width = 0.45 \textwidth]{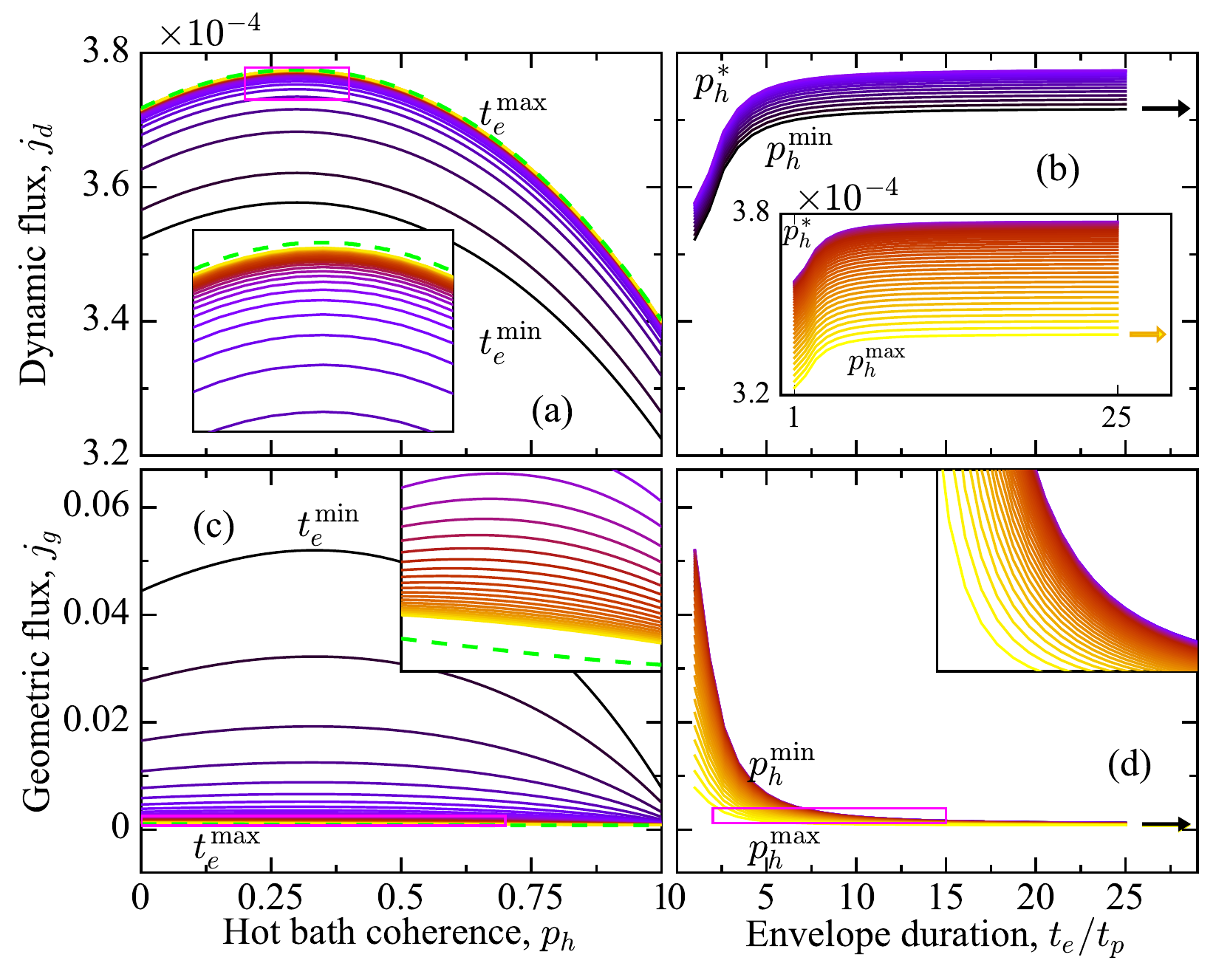}
    \caption{Graphical representation of the dynamic (panel (a) and (b)) and geometric (panel (c) and (d)) fluxes as a function of $p_{h}$ (left) for different $t_{e}$ values ($t_{e}^{\rm min}=t_{p},\,t_{e}^{\rm max}=25t_{p}$) and as a function of $t_{e}$ (right) for different $p_{h}$ values ($p_{h}^{\rm min}=0,\,p_{h}^{\rm max}=1$) for a Gaussian envelope $A_{G}(t)$. 
    (a) Dynamic flux $j_{d}$ is optimized with $p_{h}$ at various $t_{e}$.  
    $t_{e}$ increases from bottom to top.
    (b) $j_{d}$ as a function of a dimensionless envelope duration, $t_{e}/t_{p}$ (see text for interpretation). 
    (c) Optimization of geometric flux $j_{g}$ at finite envelope widths.
    $t_{e}$ decreases from bottom to top. 
    Inset shows dashed line where optimization is not possible for the sinusoidal driving. 
    (d) Decrease in $j_{g}$ with $t_{e}$.
    Arrows in (b) and (d) indicate fluxes for a sinusoidal driving for maximum (yellow) and minimum (black) $p_{h}$ values and $p_{h}^{*}$ is the optimized $p_{h}$ for dynamic flux.  
    Here: $T_{c0}=1.0,\,T_{h0}=1.67,t_{l}=2,\,E_1=E_2=0.1,\,E_b=0.4,\,E_a=1.5,\,A_0=0.01,\,\omega=2500,\,p_{c}=0.3,\,r=0.1,\,g=40,\,\tau=0.01,\,\phi=\pi/2$. 
    Atomic units are considered throughout.}
   \label{flux_fig}
\end{figure} 

\begin{figure}
    \centering
    \includegraphics[width = 0.45 \textwidth]{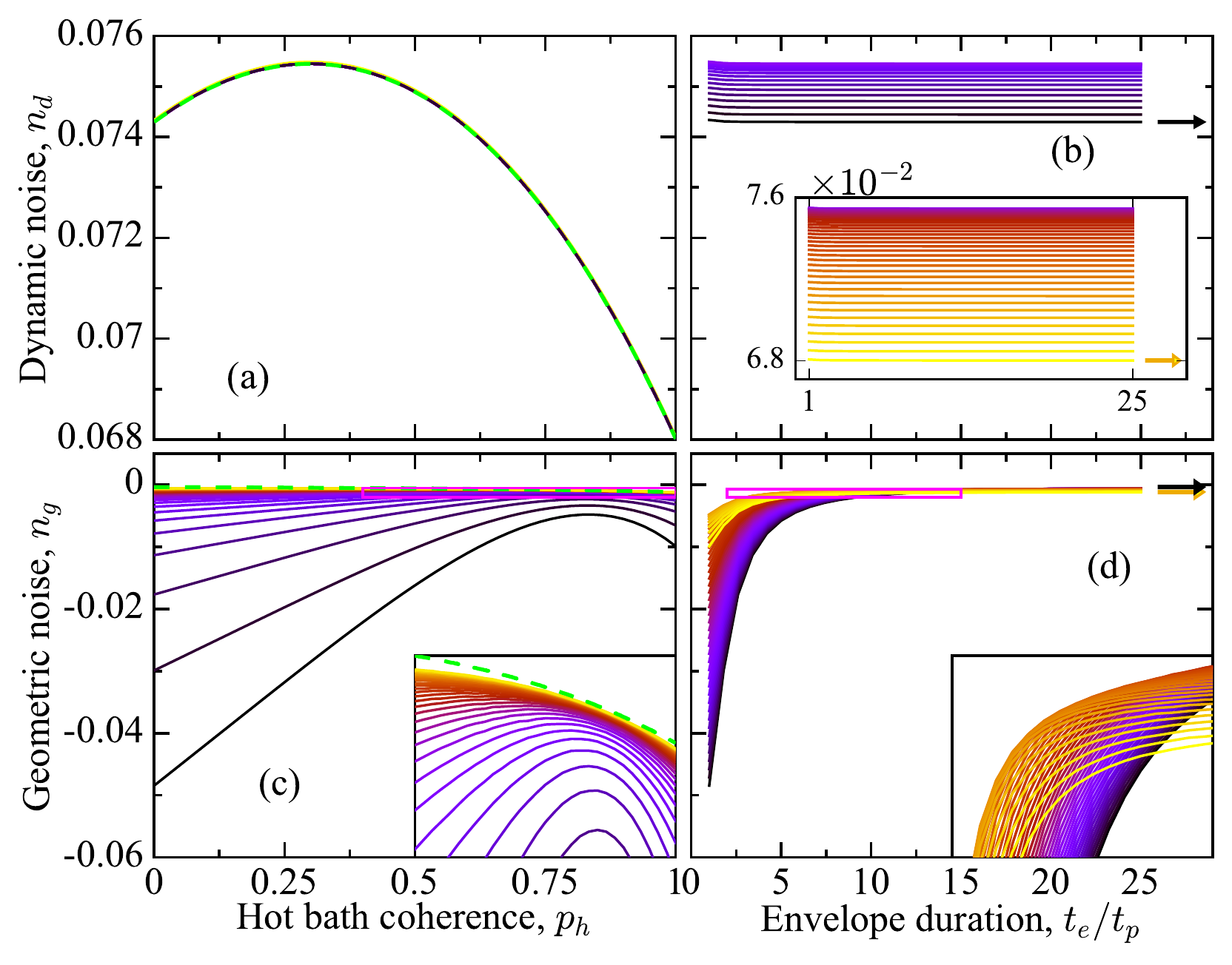}
    \caption{Dynamic ((a), and (b)) and geometric ((c), and (d)) noise for the driving with Gaussian envelope $A_{G}(t)$.
    Same color code is used as in Fig.\ref{flux_fig}. 
    Green dashed lines are for $A_{S}(t)$ envelope. 
    (a) Optimization of the dynamic noise $n_{d}$ as a function of hot bath coherence $p_{h}$ for different values of $t_{e}$. 
    Note that, the envelope shape and duration has no effect on the optimization of the dynamic noise. 
    (b) $n_{d}$ as a function of envelope duration $t_{e}$ for different values of $p_{h}$. 
    From blue to yellow $p_{h}$ increases from $p_{h}\,=\,p_{h}^{*}$ to $p_{h}\,=\,1$ and in the inset from black to blue $p_{h}$ increases from $p_{h}\,=\,0$ to $p_{h}\,=\,p_{h}^{*}$.
    (c) Optimization of the geometric noise $n_{g}$ as a function of $p_{h}$ for different values of $t_{e}$. 
    (d) Behavior of $n_{g}$ with $t_{e}$ for different values of $p_{h}$.
    In (b) and (d), black and yellow arrows represent the noise for $A_{S}(t)$ envelope for $p_{h}=0$ and $p_{h}=1$ respectively.}
   \label{noise_fig}
\end{figure}

The net photon flux exchanged between the engine and cavity is a fluctuating quantity. 
Both the flux ($j$) and the noise or fluctuations ($n$) in photon exchange are measurable quantities and are composed of additive dynamic (subscript $d$) and geometric parts (subscript $g$) given by \cite{PhysRevE.99.022104},
\begin{eqnarray}
    \label{flux_eq}
     j&\=&j_{d}+j_{g}, \\
     \label{noise_eq}
     n&=&n_{d}+n_{g}.
\end{eqnarray}
The quantities, $j_{d}\,(j_{g})$ are the first order dynamic (geometric) cumulants and $n_{d}(n_{g})$ are the second order dynamic (geometric) cumulants which can be obtained directly from a cumulant generating function described in the appendix (Eq.{\ref{s_dyn_eq} and Eq.\ref{s_geo_eq}}).  
Fig.\ref{flux_fig} and Fig.\ref{noise_fig} display the behavior of the flux and the noise respectively for the Gaussian envelope $A_{G}(t)$ (Eq.{\ref{gaussian_env_eq}}) as a function of the hot bath induced coherence $p_{h}$ and envelope duration $t_{e}$ in the unit of driving period $t_{p}$. 
It is interesting to notice the two extremum limit of the envelope $A_{G}(t)$: (i) It becomes a Dirac-delta function $A_0\delta(t)$ when $t_{e}\to 0$ and (ii) In the opposite limit, $t_{e}\to \infty$, it becomes $A_{S}(t)$ (Eq.\ref{sin_env_eq}) i.e., a sinusoidal driving. \\

In Fig.(\ref{flux_fig}a), $j_{d}$ is plotted as a function of $p_{h}$ for increasing $t_{e}$ (bottom to top) and in Fig.(\ref{flux_fig}b) $j_{d}$ is plotted against $t_{e}$ for the range $0\le p_{h}\le 1$.
For all $t_{e}$ values, $j_{d}$ is optimizable with $p_{h}$ and the optimized $p_{h}$ ($p_{h}^{*}$) is independent of $t_{e}$.
$j_{d}$ increases with $t_{e}$ rapidly and then saturates to the sinusoidal driving (green dashed line in Fig.(\ref{flux_fig}a) and arrows for two different $p_{h}$ values in Fig.(\ref{flux_fig}b)).
In Fig.(\ref{flux_fig}b), one clearly sees that the saturation threshold (the minimum value of $t_{e}$ for the saturation) does not depend on $p_{h}$. \\

In Fig.(\ref{flux_fig}c) and Fig.(\ref{flux_fig}d), the geometric flux $j_{g}$ is evaluated for the full range of $p_{h}$ and $t_{e}$. 
As a function of $p_{h}$, $j_{g}$ shows a remarkably different behavior than $j_{d}$, where we see optimization of the flux when $t_{e}$ is smaller than a critical value.
This is in contrast to what we observed earlier for sinusoidal driving, where we reported that optimization was not possible in case of  $j_{g}$ as a function of $p_{h}$ \cite{PhysRevE.96.052129}. 
But upon envelope modulation, the optimization is possible below a critical $t_{e}$ value. 
Contrary to the dynamic flux optimization, the optimal value of hot bath induced coherence, $p_{h}^{*}$, at which we see the optimized geometric flux, is dependent on $t_{e}$. 
Further, $j_{g}$ decreases as $t_{e}$ increases and eventually approaches sinusoidal driving (green dashed line in Fig.(\ref{flux_fig}c) and arrows for the two different $p_{h}$ values in Fig.(\ref{flux_fig}d)) which is complementary to the behavior of $j_{d}$.  \\

The dynamic ($n_{d}$) and geometric ($n_{g}$) noise are displayed in Fig.(\ref{noise_fig}) spanning the full range of $p_{h}$ and $t_{p}$. 
In Fig.(\ref{noise_fig}a), we have showed that $n_{d}$ is optimizable as a function of $p_{h}$ for all values of $t_{e}$. 
Interestingly, we observe that the optimization of flux and noise occurs at the same value of $p_{h}^{*}$, $p_{h}^{*}\,=\, 0.3$ for the considered parameters.    
Further, the noise does not change with $t_{e}$ and remains constant as the sinusoidal driving (green dashed line in Fig.(\ref{noise_fig}a) and arrows in Fig.(\ref{noise_fig}b)). 
This behavior is also reflected in Fig.(\ref{noise_fig}b), where $n_{d}$ for all values of $p_{h}$ is shown to be independent of $t_{e}$. 
In Fig.(\ref{noise_fig}c), the geometric noise is calculated with respect to $p_{h}$ ($t_{e}$ increases from bottom to top). 
As $t_{e}$ is decreased, $n_g$ starts exhibiting optimizable character. 
This behavior is similar to that of $j_{g}$. 
The difference is that, where $j_{g}$ decreases, $n_{d}$ increases with $t_{e}$, as shown in Fig.(\ref{noise_fig}d). 
In Fig.(\ref{noise_fig}d), we see that $n_{g}$ sharply increases at lower values of $t_{e}$ and saturates to the value obtained from sinusoidal driving (green dashed line in Fig.(\ref{noise_fig}c) and arrows in Fig.(\ref{noise_fig}d)) at different evaluated values of $p_{h}$.  \\

\begin{figure}
    \centering
    \includegraphics[width = 0.45 \textwidth]{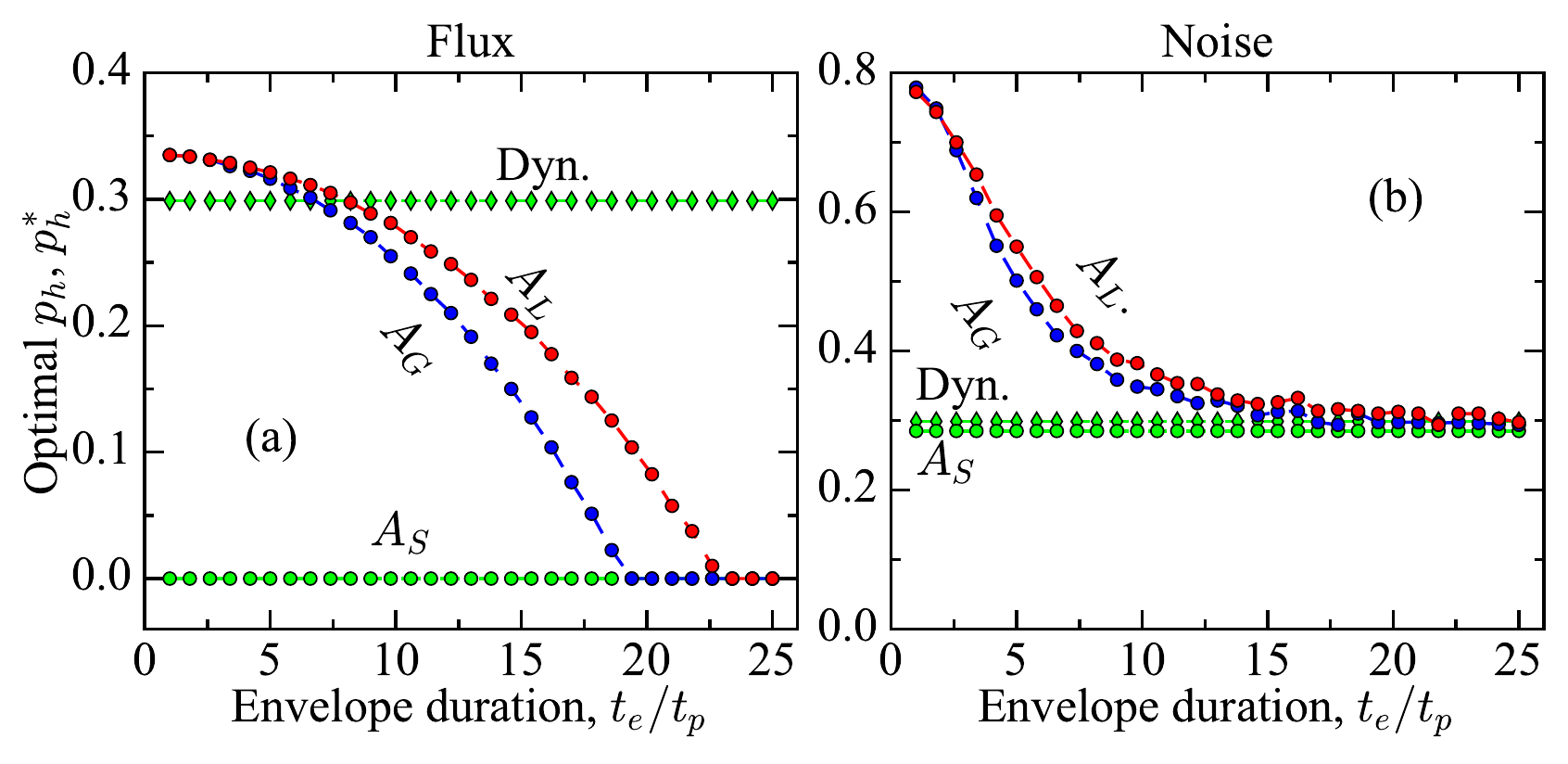}
    \caption{Variation of the optimal value of hot bath coherence $p_{h}^{*}$, as a function of envelope duration $t_{e}$ in the unit of $t_{p}$ for (a) flux and (b) noise.
    Total (dynamic + geometric) contributions are plotted along with the dynamic contribution (green diamond points) for three different envelops -- sinusoidal (green circles, Eq.\ref{sin_env_eq}), Gaussian (blue circles, Eq.\ref{gaussian_env_eq}), and Lorentzian (red circles, Eq.\ref{lor_env_eq})).
    Dynamic contribution does not depend on the envelope shape.}
   \label{phmax_fig}
\end{figure}

In Fig.(\ref{phmax_fig}), we show the dependence of the envelope duration $t_{e}$ on the optimal coherence $p_{h}^{*}$ for the dynamic and total (dynamic + geometric) flux ($j$) as well as noise ($n$). 
In Fig.(\ref{phmax_fig}a), the line at $p_{h}^{*}=0.3$
(green diamonds) represents the $j_{d}$ value and highlights the independence of $p_{h}^{*}$ on the envelope shape and duration $t_{e}$. 
The line at $p_{h}^{*}=0$ (green circles) represents the total flux when there is a sinusoidal driving. 
In the latter case, the total flux is dominated by the geometric contribution and the optimized values of flux occur at $p_{h}=0$, as known previously \cite{PhysRevE.96.052129}. 
In presence of modulated drivings (both Gaussian and Lorentzian), unlike the sinusoidal driving, the value of $p_{h}^{*}$ smoothly decreases from a large value as we keep increasing $t_{e}$ depending on the shape of the envelope. 
In Fig.(\ref{phmax_fig}b), for the dynamic noise (green diamonds) the optimal value of $p_{h}^{*}$ is independent of envelope shape and duration and also $p_{h}^{*}$ does not depend on $t_{e}$ for the sinusoidal driving (similar to what was observed for the flux). 
$p_{h}^*$ value is however larger in presence of amplitude modulation (Lorentzian or Gaussian) and gradually meets the sinusoidal driving as $t_{e}$ increases again depending on the shape of the envelope. 

\begin{figure}
    \centering
    \includegraphics[width = 0.45 \textwidth]{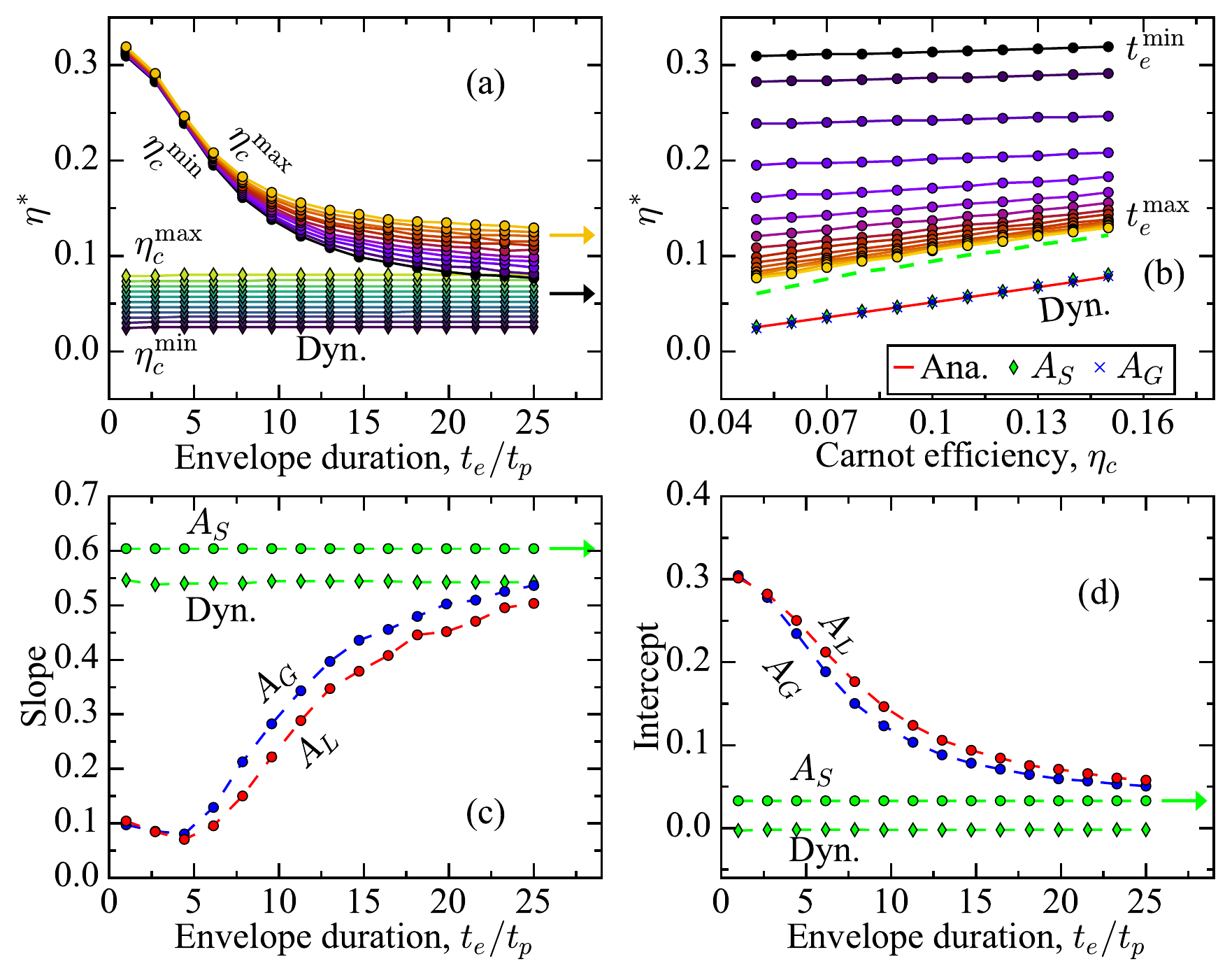}
    \caption{(a) Efficiency at maximum power $\eta^{*}$ for dynamic (blue to yellow lines) and total (black to orange lines) contributions as a function of envelope duration $t_{e}$ for increasing Carnot efficiency $\eta_{c}$ from bottom to top ($\eta_{c}^{\rm min}=0.04$ and $\eta_{c}^{\rm max}=0.15$). 
    Black and yellow arrows represent the values of $\eta^{*}$ for $A_{S}(t)$ envelope (Eq.\ref{sin_env_eq}) for minimum and maximum values of considered $\eta_{c}$ respectively.
    (b) $\eta^{*}$ as function of $\eta_{c}$ for different values of $t_{e}$ (black to yellow lines $t_{e}$ increases).
    Green dashed line is for $A_{S}(t)$ envelope (Eq.\ref{sin_env_eq}). 
    Red line is obeying the standard equation $\eta^{*}=\eta_{c}/2$, Eq.\ref{ca_eq}. 
    Green diamond and blue cross points represent dynamic part of $A_{S}(t)$ (Eq.\ref{sin_env_eq}) and $A_{G}(t)$ (Eq.\ref{gaussian_env_eq}) envelopes respectively. 
    (c) Slopes for the lines in panel (b).
    Green diamond points represent the dynamic contribution (does not depend on the envelope shape). 
    Green, blue, and red circles are for total (dynamic + geometric) contributions for $A_{S}(t)$ (Eq.\ref{sin_env_eq}), $A_{G}(t)$ (Eq.\ref{gaussian_env_eq}), and $A_{L}(t)$ (Eq.\ref{lor_env_eq}) envelopes respectively. 
    (d) Same as panel (c) but intercepts for the lines in panel (b).}
   \label{emp_fig}
\end{figure}

\begin{figure*}
\centering
\includegraphics[width = 0.7 \textwidth]{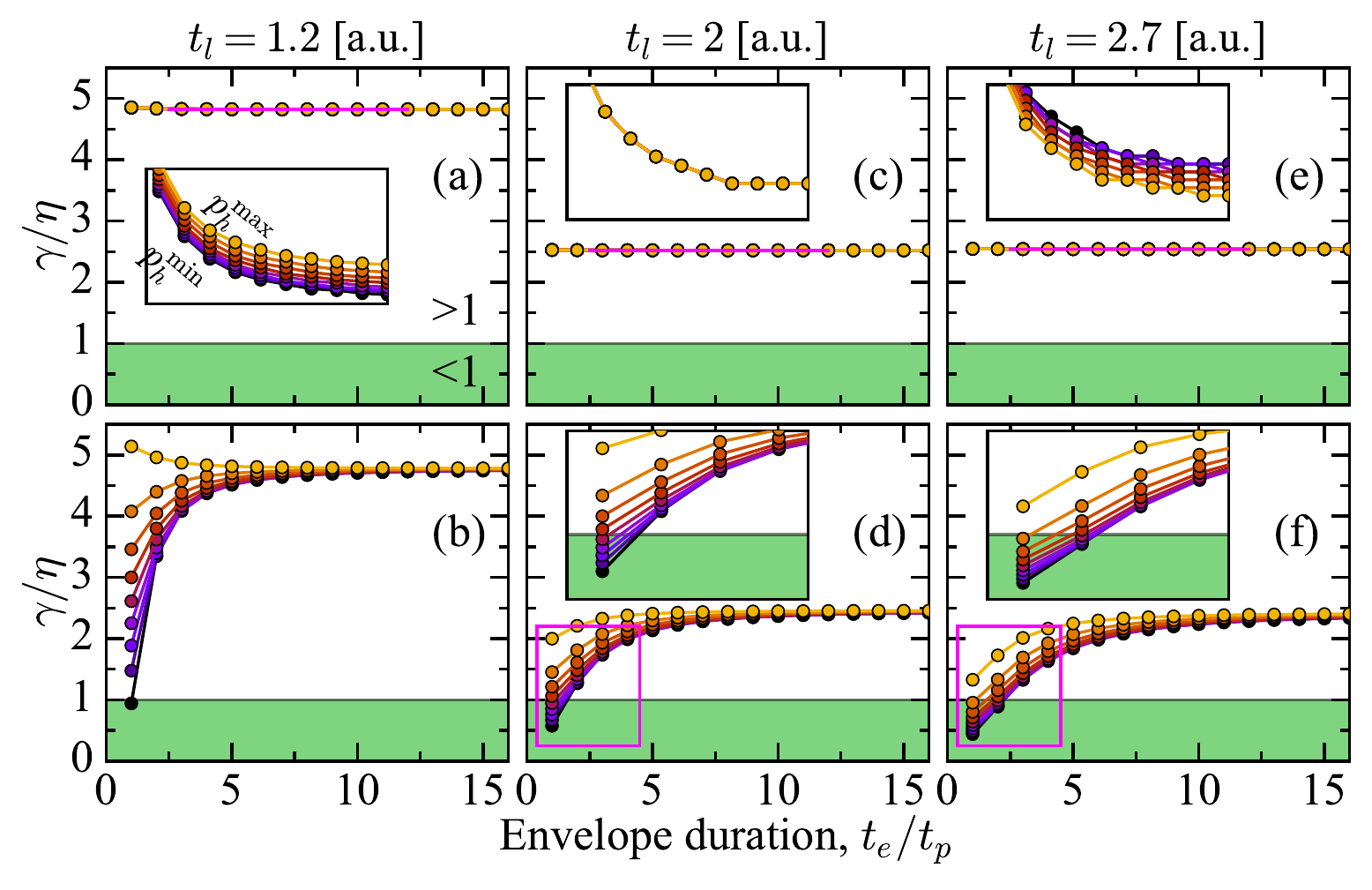}
\caption{$\gamma/\eta$ for dynamic (top) and total (bottom) contributions as a function of envelope duration $t_{e}$ for (a, b) $t_{l}=1.2$, (c, d) $t_{l}=2$, and (e, f) $t_{l}=2.7$, in the range $0\,\le\,p_{h}\,\le\,1$.}
   \label{ineq_fig}
\end{figure*}

\subsection{Efficiency and Uncertainty Relationship}
The work done by the engine is the stimulated emission of photons into a unimodal cavity coupled to the higher energy states of the engine which is given by \cite{PhysRevA.88.013842},
\begin{equation}
    \label{work_eq}
    W\=E_{a} - E_{b} +\frac{1}{t_{p}}\ln\frac{(1+n_{l})}{n_{l}}\int_{0}^{t_{p}}dt'T_{c}(t').
\end{equation}
$n_{l}$ is the cavity Bose-Einstein occupation factor expressed as $n_{l}=1/(\exp[(E_{a}-E_{b})/t_{l}]-1)$ with $t_{l}$ being the temperature of the cavity \cite{PhysRevA.88.013842}. 
The power can be expressed as, 
\begin{equation}
    \label{power_eq}
    P\=W(j_{d}+j_{g})=Wj.
\end{equation}
The efficiency of the system can be written as $\eta=W/(E_{a}-E_{1})$ and the efficiency at maximum power, $\eta^{*}$, can be obtained by optimizing $\eta$ with respect to an engine parameter (here we choose $E_b$).
A popular analytical expression for $\eta^*$, called the Curzon-Ahlborn efficiency at maximum power, can be written in terms of the Carnot efficiency, $\eta_{c}$, given by 
\begin{eqnarray}
     \eta^{*}&\=&1-\sqrt{1-\eta_{c}}\\
     \label{ca_eq}
     &\=&\frac{\eta_{c}}{2}, ~ \text{near equilibrium.}
\end{eqnarray}
The linear coefficient, $1/2$ has been claimed to be universal \cite{PhysRevLett.95.190602, PhysRevLett.102.130602}, which we showed was violated in presence of geometric effects\cite{PhysRevE.96.052129} with sinusoidal drivings.  
Here, we evaluate $\eta^{*}$ as a function of the envelope duration $t_{e}$ of the modulated driving and compare the results with Eq.(\ref{ca_eq}). 
In Fig.(\ref{emp_fig}a), we show the behavior of $\eta^{*}$ with respect to $t_{e}$ for the range $0.05\,<\,\eta_{c}\,<\,0.15$ (bottom to top). 
The total $\eta^{*}$ is maximum ($>\,0.3$) when the envelope duration is minimum $(t_{e}=t_{p})$.  $\eta^{*}$ non-linearly decreases as $t_{e}$ increases and eventually saturates to the value obtained from sinusoidal drivings at large $t_{e}$. 
The lower set of curves (parallel lines) correspond to $\eta^{*}$ values, when there are no geometric contributions (only dynamic). Here, $\eta^{*}$ does not depend on $t_{e}$. 
In Fig.(\ref{emp_fig}b), we show that $\eta^{*}$ linearly increases with $\eta_{c}$, but the slope is $1/2$ only when geometric contributions are absent (holds for dynamic, green diamond and blue cross points). 
Further, $t_{e}$ has no effect on the slope under the same conditions. 
However, in presence of geometric effects, this slope of $1/2$ is not maintained anymore. 
The behavior of the slope and intercept in presence of geometric effects is shown graphically in Figs.(\ref{emp_fig} c and d) respectively.
The slope decreases, reaches a minimum, and then gradually increases and saturates at the respective values obtained for sinusoidal case, for both Gaussian and Lorentzian drivings. 
Note that, as per Eq.(\ref{ca_eq}), there is no intercept in $\eta^{*}$ as a function of $\eta_{c
}$. 
In presence of geometric effects, an intercept is introduced in the standard expression because of the driven dynamics. 
This intercept non-linearly decreases with $t_{e}$ and approaches the value obtained from sinusoidal drivings. \\

Efficiency, being one of the most characteristic quantity of engines is often deeply investigated to gain deeper thermodynamic insights. 
During the last two years, with respect to QHE, several interesting bounds on efficiency have been proposed \cite{NatPhy.16.15}, especially derived from TUR \cite{NatPhys.16.15}. 
One of such bounds on the efficiency of QHEs is given by \cite{PhysRevLett.126.210603},
\begin{equation}
    \label{miller_tur_eq}
    \gamma/\eta\,\ge\,1, ~~ \text{with}~ \gamma\=\frac{\eta_{c} P}{T_{c}\dot\Sigma+P},
\end{equation}
where $\dot\Sigma$ represents the rate of entropy production and has been claimed to be a direct result of TUR in quantum systems \cite{PhysRevLett.125.260604}. 
The average entropy production is given by $\dot\Sigma = j {\cal A}$, where ${\cal A}$ is the thermodynamic affinity \cite{PhysRevB.98.155438}. 
Using an established TUR of the type ${\cal A}n/j\,\ge\,2k_{B}$\cite{PhysRevB.98.155438}, it is straight-forward to recast
Eq.(\ref{miller_tur_eq}) to ($k_{B}\to 1$),
\begin{equation}
     \label{eta_bound_eq}
     \gamma/\eta\,\ge\,1, ~~ \text{with} ~ \gamma\=\frac{\eta_{c} P}{P+T_{c}{\cal A}j}.
\end{equation}
 It is natural to see the validity of the Eq.(\ref{eta_bound_eq}) in presence of geometric effects as well as other engine parameters. 
 In our engine, the thermodynamic affinity is known and is given by\cite{PhysRevE.99.022104}
 ${\cal A}\,=\,\ln\frac{\widetilde{n}_{l}\int_{0}^{t_{p}} dt' [1+n_{c}(t{'})]n_{h}(t{'})}{n_{l}\int_{0}^{t_{p}} dt'n_{c}(t{'})[1+n_{h}(t{'})]}$. 
 We numerically evaluate $\gamma$ and $\eta$ in Eq.(\ref{eta_bound_eq}) and plot $\gamma/\eta$ as a function of envelope duration, $t_{e}$ in Fig.(\ref{ineq_fig}), evaluated at different cavity temperatures $t_{l}$ and coherence values $p_{h}$. 
 In the panels (a), (c), and (e), there are no geometric contributions (only dynamic, $\phi\,=\,0$) and the inequality $\gamma/\eta\,>\,1$ is always maintained irrespective of any engine parameters.
From the insets, we show that $\gamma/\eta$ changes its order with respect to $p_{h}$ as $t_{l}$ increases (panel (a) to (e)).  
Most interestingly, in the presence of geometric contribution (panel (b), (d), and (f)), by suitably selecting $t_{l}$ and $t_{e}$ we report  a region where the inequality Eq.\ref{eta_bound_eq} does not hold. 
This happens at very small values of $t_{e}$ and large values of $t_{l}$ where we observe that $\gamma/\eta\,<\,1$.
As $t_{e}$ increases and the driving approaches the value obtained from sinusoidal drivings where the inequality is recovered.
Therefore, the inequality condition is broken only in the presence geometric effects introduced due to amplitude modulation. 
If the amplitude modulation is absent, the inequality holds. \\

\section{Conclusion}
\label{conc}
In this work, we chose to drive the two temperatures of the thermal reservoirs of a quantum heat engine with protocols where the driving amplitude is being modulated in the adiabatic limit introducing envelope functions. 
With such amplitude modulation, we reported the optimization of the geometric flux with respect to quantum coherences for a finite envelope duration, which is otherwise not possible with simple sinusoidal drivings. 
Further, we also optimized the dynamic as well as the geometric noise and this optimization is independent of envelope duration for the former one whereas for the later one optimization point is envelope duration dependent. 
The optimal value of coherence decreases as the envelope duration is increased depending on the shape of envelope. 
Another interesting thermodynamics quantity, the efficiency at maximum power (EMP), decreases non-linearly with the envelope duration. 
In the presence of both geometric effects and modulated driving with envelope, the slope and intercept arises, which deviate from the standard linear expression for EMP in terms of Carnot efficiency in an intricate manner depending on the shape and duration of the envelope. 
Further, universal bounds on efficiency based on uncertainty relationships does not hold when geometric effects are employed via amplitude modulation with shorter envelope duration and larger cavity temperatures.

\section*{Appendix}
The QHE has degenerate quantum states $|1\rangle$ and $|2\rangle$, with same symmetry (therefore with a forbidden transition between them) are coupled to two thermal baths. 
The higher energy states $|a\rangle$ and $|b\rangle$ with different symmetry and allowed transition between them are coupled to the hot and cold bath respectively. 
The state $|a\rangle$ is higher in energy than the state $|b\rangle$. 
$|1\rangle$, $|2\rangle$, $|b\rangle$ and $|a\rangle$ states correspond to the energies of $E_1$, $E_2$, $E_b$ and $E_a$ respectively. 
States $|a\rangle$ and $|b\rangle$ are also coupled to a unimodal cavity and the strength of the coupling is denoted by $g$. 
With above assumptions the total Hamiltonian can be written as $\hat{H}_{T}\,=\,\hat{H}_{0}+\hat{V}_{sb}+\hat V_{sc}$, where
\begin{eqnarray}
    \label{hamiltonian_eq}
    \begin{split}
        \hat{H}_{0}&\=\sum_{\nu\,=\,1,2,a,b}
        E_{\nu}|\nu \rangle\langle \nu |+\displaystyle\sum_{k\in h,c}\epsilon_{k}\hat{a}_{k}^{\dag}\hat{a}
        _k+\epsilon_l\hat{a}_{l}^{\dag}\hat{a}_{l}, \\
        \hat V_{sb}&\=\sum_{k\,\in\,h,c}\sum_{i\,=\,1,2}\sum_{x\,=\,a,b}r_{ik}\hat{a}_{k}|x\rangle\langle i|+\text{H.c},  \\
        \hat V_{sc}&\=g\hat{a}_{l}^\dag|b\rangle\langle a|+\text{H.c}.
    \end{split}
\end{eqnarray} 
In the above equation, $E_{\nu}$, $\epsilon_{k}$ and $\epsilon_{l}$ are the energy of the system's $\nu$th level, $k$th mode of the thermal reservoirs and unimodal cavity respectively.
$r_{ik}$ is the system-reservoir coupling of the $i$th state with the $k$th mode of the reservoirs. 
Thermal baths are modeled as harmonic modes with $\hat{a}^\dag (\hat{a})$ being the bosonic creation (annihilation) operators. 
There is a heat flow from the hot bath to the cold bath in a nonlinear fashion. 
Also, there is a radiative decay channel originates from the transition $|a\rangle\to|b\rangle$. \\

Apropos to the theoretical formalism in the Liouville space, presented in our earlier works \cite{PhysRevE.96.052129, PhysRevE.99.022104},  a reduced density vector in the Liouville space  is composed of the four coupled population and a coherence given by $|\rho(\lambda,t)\rangle = \{\rho_{11},\rho_{22},\rho_{aa},\rho_{bb},\Re(\rho_{12})\}$, with $i=1,2,a,b$ which denotes the system's many body states and $\Re(\rho_{12})$ is the thermally induced coherence between states $|1\rangle$ and $|2\rangle$.  
An adiabatic Markovian quantum master equation approach combined with a standard generating function technique allows us to evaluate the statistics of photons exchanged between the engine and cavity as per the equation $\dot\rho(\lambda,t)\rangle = \breve{{\cal L}}(\lambda,t)|\rho(\lambda,t)\rangle$, where  $\lambda$ is a field that counts the number of photons exchanged between the system and the cavity. 
$\breve{{\cal L}}(\lambda,t)$ is the adiabatic effective evolution Liouvillian superoperator within the Markov approximation, given by
\begin{widetext}
    \begin{equation}
        \label{liouvillian_lambda_eq}
        \breve{{\cal L}}(\lambda,\,t)\= 
        \begin{pmatrix}
            n_{1}(t)&0&r_{1h}\tilde{n}_{h}(t)&r_{1c}\tilde{n}_{c}(t)&y(t) \\
            0&n_{2}(t)&r_{2h}\tilde{n}_{h}(t)&r_{2c}\tilde{n}_{c}(t)&y(t) \\
            r_{1h}n_{h}(t)&r_{2h}n_{h}(t)&-g^{2}\tilde{n}_{l}-2r_{h}\tilde{n}_{h}(t)&g^{2}n_{l}\e^{-\lambda}&2r_{h}p_{h}n_{h}(t) \\
            r_{1c}n_{c}(t)&r_{2c}n_{c}(t)&g^{2}\tilde{n_{l}}\e^{\lambda}&-g^{2}n_{l}-2r_{c}\tilde{n}_{c}(t)&2r_{c}p_{c}n_{c}(t) \\
            \frac{y(t)}{2}&\frac{y(t)}{2}&r_{h}p_{h}\tilde{n}_{h}(t)&r_{c}p_{c}\tilde{n}_{c}(t)&-n(t) 
        \end{pmatrix}
    \end{equation}
\end{widetext}
In the above equation $n_{i}(t)\,=\,-[r_{ic}n_{c}(t)+r_{ih}n_{h}(t)]$
with $i\,=\,1,2$, $r_{c}\,=\,r_{1c}+r_{2c}$, $r_{h}\,=\,r_{1h}+r_{2h}$,
$y(t)\,=\,-r_{c}n_{c}(t)p_{c}-r_{h}n_{h}(t)p_{h}$, 
$\tilde{n}_{c}(t)\,=\,n_{c}(t)+1$, 
$n(t)\,=\,(r_{1h}+r_{2h})n_{h}(t)/2+(r_{1c}+r_{2c})n_{c}(t)/2+\tau$, 
$\tilde{n}_{h}(t)\,=\,n_{h}(t)+1$, 
$\tilde n_{l}=1+n_{l}$ and $\tau$ is an environmental dephasing parameter. 
In this study we have considered equal system-reservoir coupling denoted by $r$.
The explicit form of $n_{c}(t)$ and $n_{h}(t)$ can be expressed as $ n_{c}(t)\,=\,1/(\exp\{(E_{b}-E_{1})/k_{B}T_{c}(t)\}-1), n_{h}(t)\,=\,1/(\exp\{(E_{a}-E_{1})/k_{B}T_{h}(t)\}-1)$.  $p_{h}$ and $p_{c}$ is the quantum coherence control parameters associated with the hot and cold baths respectively.\\

The statistics of $q$ (number of photons exchanged between the system and the  cavity) is obtained from  moment generating function, which is expressed as $G(\lambda,t)\,=\,\sum_q\e^{\lambda q}P(q,t)$ where $P(q,t)$ is  the probability distribution function corresponding to $q$ net photons in the cavity  within a measurement window, $t$. 
Within the full counting statistics (FCS) formalism, it can be shown that $\dot G(\lambda,t)\,=\,\langle \breve{{\boldsymbol 1}}| \breve{\mathcal L}(\lambda,t)|\rho(\lambda,t)\rangle$ with $\langle\breve{\bf 1}|\,=\,\{1,1,1,1,0\}$ \cite{PhysRevB.70.115305, RevModPhys.81.1665}. With the help of Eq.(\ref{liouvillian_lambda_eq}), one can obtain geometric contributions from the scaled cumulant generating function given by $S(\lambda)\,=\,\lim_{t\rightarrow\infty}(1/t)\ln[\langle\breve{\bf 1}|\exp(\breve{\cal L}(\lambda,t)t)|\rho(\lambda,t)\rangle]$. $S(\lambda)$ is separable into dynamic and geometric parts additively, $S(\lambda,t)\,=\,S_d(\lambda,t)+S_g(\lambda,t)$,
 \begin{eqnarray}
     \label{s_dyn_eq}
     S_{d}(\lambda)&\=&\frac{1}{t_{p}}\int_{0}^{t_{p}}dt'\zeta_o(\lambda,t'),\\
     \label{s_geo_eq}
     S_{g}(\lambda)&\=&-\frac{1}{t_{p}}\int_{0}^{t_{p}}\langle L_{o}({\lambda,t})
     |\dot R_o({\lambda,t})\rangle dt.
 \end{eqnarray}
In the above equation, $S_{d}(\lambda)$ and $S_{g}(\lambda)$ represent the dynamic and geometric cumulant generating function respectively. $|R_{o}(\lambda,t)\rangle$ and $\langle L_{o}(\lambda,t)|$ are the instantaneous right and left eigenvectors of $\breve{\cal L}(\lambda,t)$ with instantaneous long-time dominating eigenvalue, $\zeta_o(\lambda,t)$. 
Note that, analytical expressions for both $S_{d}(\lambda)$ and $S_{g}(\lambda)$ cannot be derived for $4$ level dQHE. 
The cumulant generating function are analytically known only for two level systems\cite{PhysRevLett.104.170601, PhysRevB.93.195441} within the Markov limits. Systems with large number of  states, analytical expressions have not been reported since the geometric contributions involve calculation of both the left and right eigenvectors of the Hamiltonian. 
The $n$th order fluctuations (cumulants of $S(\lambda)$) can be calculated as
\begin{eqnarray}
    \label{cd_eq}
    C_{d}^{(i)}&\=\partial_\lambda^{i} S_{d}(\lambda)|_{\lambda\,=\,0}, \\
     \label{cg_eq}
    C_{g}^{(i)}&\=\partial_\lambda^{i} S_{g}(\lambda)|_{\lambda\,=\,0}.
\end{eqnarray}
When $i\,=\,1$, we get the dynamic (geometric) flux, $j_{d}(j_{g})$, and when $i\,=\,2$, we obtain the dynamic (geometric) noise, $n_{d}(n_{g})$, which are numerically evaluated.   

\begin{acknowledgments}
  HPG acknowledges the support from Science and Engineering Board for the start-up grant, SERB/SRG/2021/001088. 
\end{acknowledgments}  

\bibliography{references.bib}

\end{document}